\documentclass[preprint,5p,twocolumn,superscriptaddress,color,showpacs]{elsarticle}

\usepackage{amsmath,amsfonts,amssymb,color,verbatim,wasysym}

\begin{document}

\begin{frontmatter}

\title{Threshold cascade dynamics on signed random networks}

\author[inst1,inst2]{Kyu-Min Lee}
\author[inst1]{Sungmin Lee}
\author[inst3]{Byungjoon Min}
\ead{bmin@cbnu.ac.kr}
\author[inst1]{K.-I.\ Goh}
\ead{kgoh@korea.ac.kr}

\address[inst1]{Department of Physics, Korea University, Seoul 02841, Korea
}
\address[inst2]{College of Business, Korea Advanced Institute of Science and Technology, Seoul 02455, Korea
}
\address[inst3]{Department of Physics, Chungbuk National University, Cheongju 28644, Korea
}

\date{\today}

\begin{abstract}
Relationships between individuals in a social network, genes in biological systems, and spins in magnetic 
systems often reflect a mixture of positive (friendly) and negative (antagonistic) interactions. 
However, most studies of complex networks have focused on networks consisting of solely positive 
interactions. Here, we study threshold cascades on signed networks composed of 
both positive and negative connections, focusing on when a pair of nodes connected by a negative link can only be activated exclusively to each other. We found that the negative interactions 
not only suppress global cascades, but also induce the heterogeneity in activation 
patterns manifesting from single-node to network levels. Our results suggest that negative interactions
may be an important source of the variability in cascading dynamics.
\end{abstract}


\end{frontmatter}

\section{Introduction}

Modeling how the cascades of activations occur in threshold-based dynamics is 
fundamental for understanding collective behaviours in social and biological complex 
systems~\cite{schelling,granovetter,watts2002,rohlf2002}. In order to model the 
cascading phenomena triggered by a tiny perturbation, a threshold cascade model 
was proposed~\cite{granovetter,watts2002}. This model was originally motivated by 
the behavioral and emotional contagions in a society where individuals are encouraged to 
follow what their connected neighbors are doing. In addition, threshold cascades 
driven by integrate-and-fire mechanisms are associated with the avalanches of neural 
activations~\cite{friedman2012,kusmierz2020}, the spread of economic 
crisis~\cite{kmlee2011}, and cascading failures in infrastructure 
networks~\cite{buldyrev2010,brummitt2012a,bmin2014}. The key mechanism in this 
model is that nodes are activated when the fraction of activated neighbors exceeds their 
threshold assigned a priori. In this model, cascades with an extensive size, called global 
cascades, can occur from an extremely small fraction of seeds because the cascades of 
activations propagate along connected neighbors~\cite{watts2002,gleeson2007}.

In threshold cascade models on networks, links act as channels for cascade propagation, such 
that the influence or stimulus arriving from each neighbor contributes positively to reaching 
the threshold \cite{watts2002}. Although traditional cascade modeling, which consists  
exclusively of positive links \cite{watts2002,gleeson2007,motter2002,hackett2011,brummitt2012b,kmlee2014}, 
renders the model simple and tractable, it overlooks the negative interactions in 
the cascade dynamics. Adversarial interactions are common and essential elements of 
many networked systems~\cite{bowers2004,szell2010,leskovec2010a,rubinov2010,tang2016}. 
``Dislike'' relationships in social networks~\cite{szell2010,leskovec2010a,tang2016,leskovec2010b,bjkim}, 
inhibitory signals in genetic regulation~\cite{bowers2004,regulondb,regulondbv10}, synaptic inhibition
in neural networks~\cite{rubinov2010}, antagonistic competitions between nations~\cite{maozbook,maoz2007}, 
and antiferromagnetic bonds in magnetic systems~\cite{parisibook} are typical examples 
of adversarial relationships, to name a few. Not only negative links are widespread in real-world
systems, but also they play a qualitatively different role in dynamical processes 
than positive links~\cite{bowers2004,rubinov2010}.
Networks with both types of interactions can be better modeled as ``signed networks'' where 
links are either positive or negative~\cite{leskovec2010a,heider1946,cartwright1956,facchetti2011,ciotti2015}. 
The concept of signed networks has long been proposed in psychology and sociology, through 
social balance theory~\cite{heider1946} and structural balance theory~\cite{cartwright1956,du2016,he2018}. 
In addition, from the perspective of statistical physics, coexistence of positive and negative 
interactions has important implications as a source of geometric frustration and dynamic 
heterogeneity~\cite{parisibook,dhkim2005}. As such, the studies on signed networks have 
received due attention from statistical physics and network science 
communities~\cite{facchetti2011,antal2005,antal2006}.
However, studies on the impact of negative interactions on threshold cascade dynamics
are still lacking.

In this work, we study the dynamics of a threshold cascade model on signed random networks
or ``signed'' cascade, to be short. 
In our ``signed'' cascade model, nodes' activation is completely blocked if there exist active adversarial neighbors. 
That is, no pair of nodes connected by a negative link can be activated at the same time. 
It models, in an idealized way, the following real-world scenarios: 
In the case of ``distrust'' or ``dislike'' relationships denoted by negative links in a signed social network, someone would never agree and follow with their negative-linked person's opinion or behavior, regardless of what their friends are doing.  
In such cases, a pair of nodes connected by a negative link cannot become active at the same time. 
Theoretically speaking, our model implements the cascade dynamics where negative links co-operate with positive links in functionally-multiplicative manner rather than additive manner.

There are some related previous studies 
in statistical physics literature on signed networks such as random threshold 
networks~\cite{rohlf2002}, percolation of antagonistic multiplex networks~\cite{zhao2013a}, opinion 
models on evolving signed networks~\cite{quattrociocchi2014}, epidemic spreading on signed 
networks~\cite{saeedian2016,li2021}, and threshold model with anticonformity~\cite{nowak2019,nowak2022}.
Information diffusion and linear threshold models in signed networks have also received much 
attention and how the information diffusion dynamics in signed networks depends on the diffusion path
and structural balance was studied \cite{he2019,pozveh2019,li2019,qu2021}.
However, in contrast to our model most previous studies have considered the effect of negative 
links additively~\cite{rohlf2002,nowak2019,he2019,li2021} or focused on structural properties rather 
than dynamical consequences~\cite{zhao2013a}.
In this study, we implemented cascading 
dynamics in a way that maximizes the multiplicative coupling effect of negative links. 
We confirmed that the negative links can significantly reduce the size of the global 
cascades. We also found that negative interactions can produce the heterogeneity 
in the activation patterns at various scales of cascading dynamics.

\section{Signed cascade model}

\begin{figure}
\centering
\includegraphics[width=\linewidth]{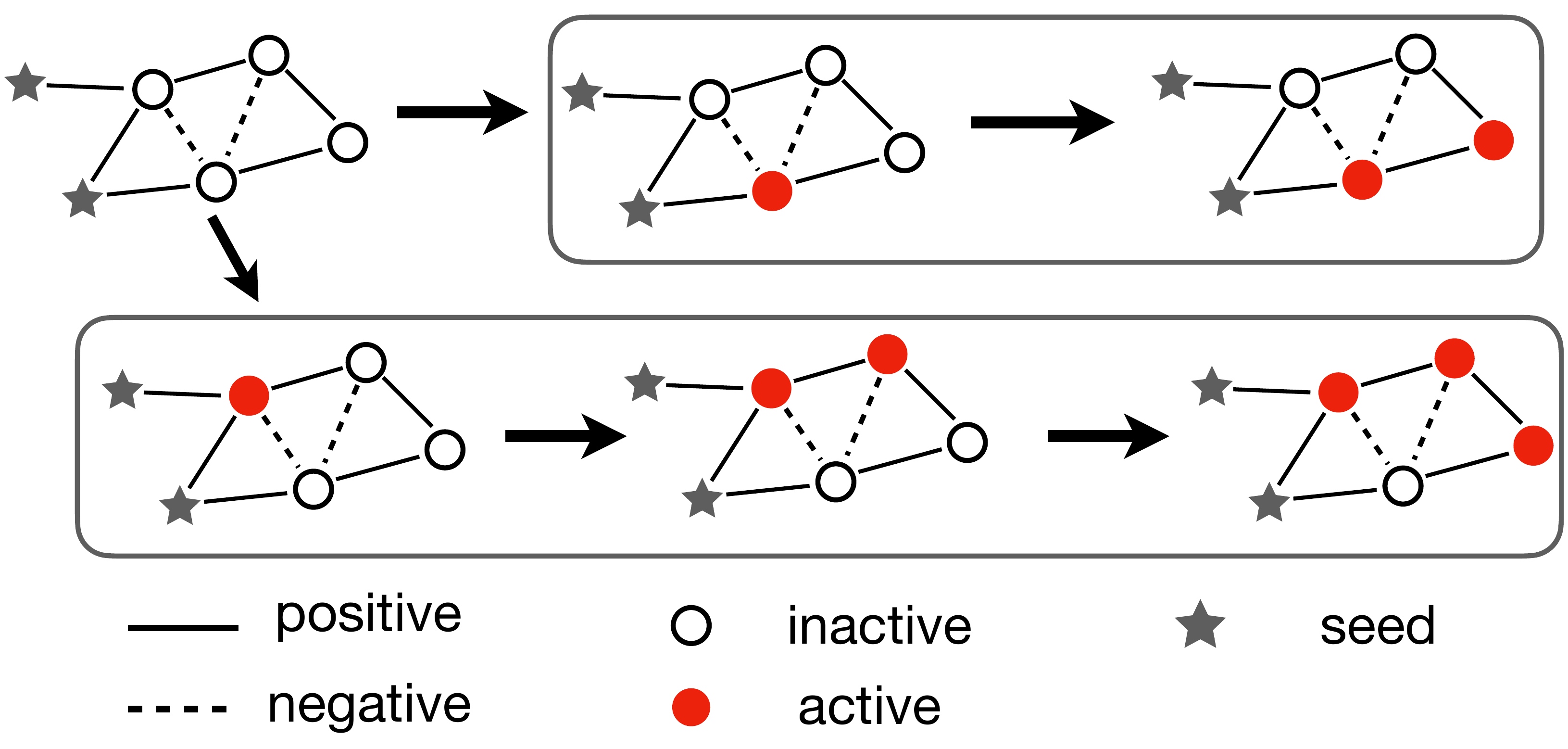}
\caption{
Illustration of threshold cascades on a signed network.
If the fraction of active neighbors of node $i$ connected by positive links is 
larger than threshold $\theta=0.4$ and there are no active neighbors connected 
by negative links, node $i$ becomes active. Note that there can be various scenarios
of cascades activations depending on the sequence of activations.
}
\label{fig:model}
\end{figure}

We propose a model of threshold cascading dynamics on a signed network by explicitly 
implementing the role of negative links preventing the activation of 
a connected neighbor.
Each node can be one of two states, active or inactive in multiplicative manner. 
In signed networks, each link
can be positive or negative. Neighbors connected by positive (negative) links 
are referred to as positive (negative) neighbors for short.
A node becomes active when the following two conditions are fulfilled:
i) the fraction of active positive neighbors out of total positive neighbors exceeds
the prescribed threshold $\theta$ as in the ordinary threshold model \cite{watts2002} 
and ii) there are no active negative neighbors. 
The rule clearly shows different roles of the positive and negative connections in the 
``signed'' cascades. While positive links spread activations to a connected neighbor,
negative links prevents neighbors from activation. 
Logical ``AND'' requirement of both conditions reflects the multiplicative coupling of 
positive and negative interactions. According to the model if there 
exists even one active negative neighbor, the activation of the node is completely 
blocked. We impose the strongest role of negative links in order to demonstrate 
the effect of the signed networks in a simple and dramatic way.

Let us describe the procedures for numerical simulations of threshold cascading on 
a signed network. Initially all nodes are inactive except for a small 
fraction $\rho_0$ of the seed nodes that are active at the beginning. The signed 
cascade proceeds as follows. i) At each step, we select a node, say $i$, 
at random. ii-a) For inactive node $i$, the state of node $i$ becomes active when 
the ratio of its active positive neighbors exceeds threshold $\theta$ and there is no active 
negative neighbor. For instance, suppose that there are $r$ active positive neighbors
out of $k_p$ positive neighbors for node $i$. Then node $i$ becomes active when 
$r/k_p>\theta$ and there is no active negative neighbor. ii-b) If an active node 
including a seed is selected, nothing happens. It means that active nodes maintain 
their active state permanently. iii) The procedures repeat until the dynamics of activations 
reaches a steady state meaning that there exists no node that can be newly activated. 
An example of the signed cascade process is depicted in Fig.~\ref{fig:model}. We perform
random sequential updates, so that we choose at random a single node in the network 
and update its state at every step. Note that if there is no negative link, our model reduces
to the original Watts threshold model \cite{watts2002}.

Contrary to the original Watts threshold model, cascading dynamics on signed 
networks is no longer deterministic because of the role of negative interactions.
Specifically, the set of active nodes in a steady state can be diverse even when 
the cascading dynamics starts from identical seeds on the same network structure.
The final configuration is stochastically realized among many possible configurations 
that satisfy the conditions of both activation and inactivation as 
illustrated in Fig.~\ref{fig:model}. 
The variability of the final configuration depending on the sequence of 
activations produces more heterogeneous and richer dynamics than 
that of the traditional threshold model.

\section{Results}

\subsection{Suppression of Global Cascades}

\begin{figure*}
\centering
\includegraphics[width=0.99\linewidth]{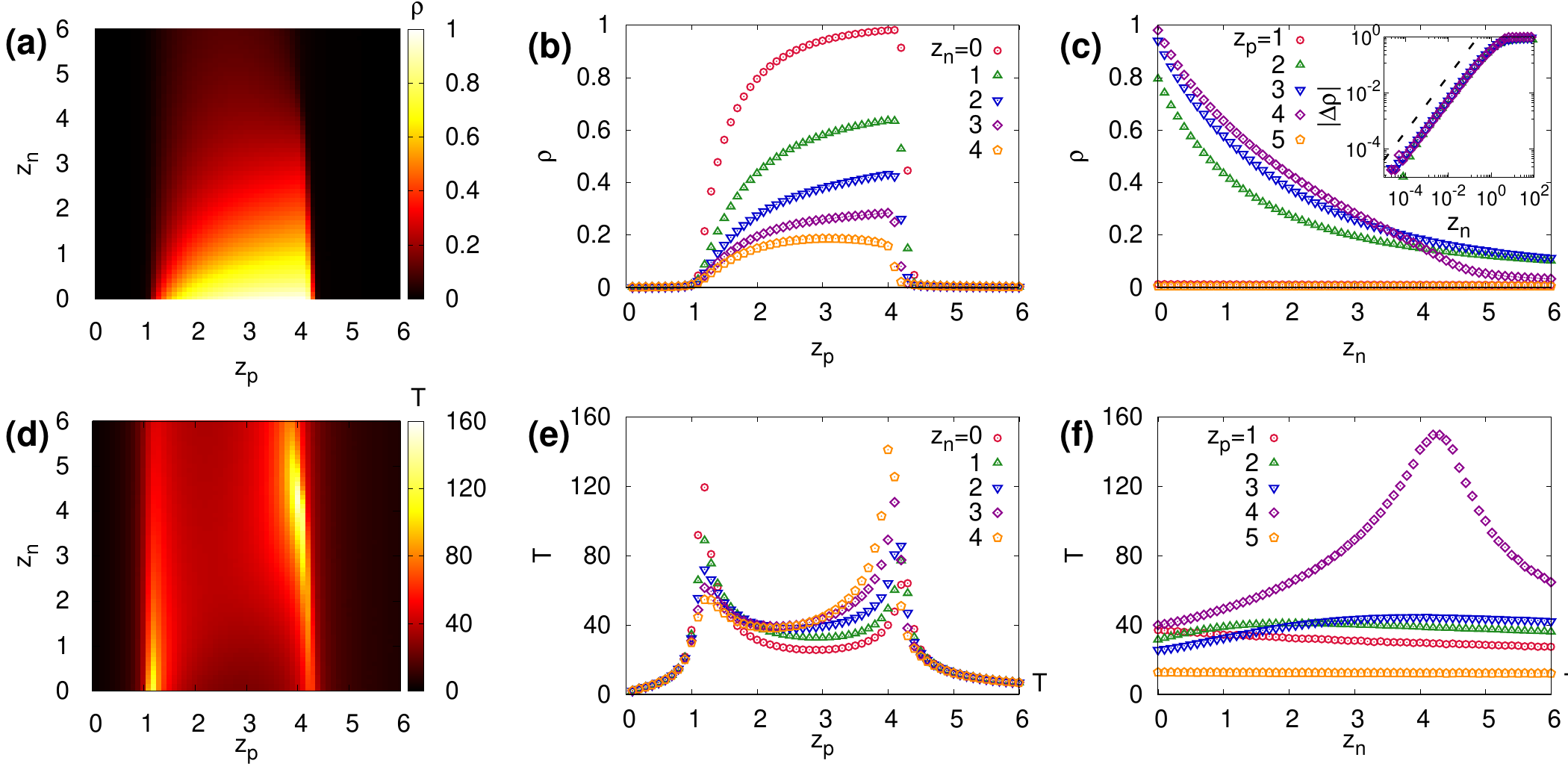}
\caption{
(a) The size $\rho$ of global cascades with respect to $z_n$ and $z_p$ on ER networks 
with $N=10^5$, seed fraction $\rho_0=10^{-3}$, and threshold $\theta=0.2$, averaged 
over $10^5$ different realizations.
(b) The size $\rho$ of global cascades as a function of $z_p$ for various $z_n=0,1,2,3,4$.
(c) The size $\rho$ of global cascades as a function of $z_n$ for various $z_p=1,2,3,4,5$.
Inset shows that the gap $\Delta \rho$ between $\rho$ at $z_n$ and $z_n=0$ increases 
linearly with increasing $z_n$ for $z_p=2,3,4$.
Numerical results (symbols) and theoretical predictions (lines) are shown together.
(d) The duration $T$ of cascading activation with respect to $z_n$ and $z_p$ on ER networks 
with $N=10^5$, seed fraction $\rho_0=10^{-3}$, and threshold $\theta=0.2$, averaged 
over $10^5$ different realizations.
(e) The duration $T$ of global cascades as a function of $z_p$ for  various  $z_n=0,1,2,3,4$.
(f) The duration $T$ of global cascades as a function of $z_n$ for  various  $z_p=1,2,3,4,5$.
}
\label{fig:rho}
\end{figure*}

The primary effect of negative links is the suppression of cascading dynamics due 
to the local suppression by antagonistic connections. We examine how negative links globally 
suppress cascading dynamics on a signed network focusing on the global cascades. 
We measured the size of global cascades as the fraction $\rho$ of active nodes 
at the steady state. We use Erd\H{o}s-R\'enyi (ER) graphs with network size $N=10^5$, 
seed fraction $\rho_0=10^{-3}$, and threshold $\theta=0.2$ for every node unless otherwise stated.
As control parameters, we vary the average degrees for positive links $z_p$ and for negative links $z_n$. 
The global cascade size $\rho$ as a function of $z_p$ and $z_n$ is numerically 
calculated [Fig.~\ref{fig:rho}(a)]. We found that the size $\rho$ of global cascades 
monotonically decreases by adding more negative links 
as shown in the vertical gradient in Fig.~\ref{fig:rho}(a) 
owing to the suppression effect of negative links.

Let us examine the behaviour of $\rho$ as a function of $z_p$ with various $z_n=0$ 
($\circ$),  $z_n=1$ ($\triangle$), $z_n=2$ ($\triangledown$), $z_n=3$ ($\diamond$), 
$z_n=4$ ($\pentagon$) in Fig.~\ref{fig:rho}(b). The global cascade can occur 
only for a region with an intermediate range of $z_p$. When $z_p$ is less than
$z_p^{(1)}$ which is the percolation threshold of random graphs \cite{molloy1998}, there 
is no global cascade in the limit $\rho_0 \rightarrow 0$ because the
components cannot span a finite fraction of a network in the thermodynamic 
limit. Therefore the effect of negative links cannot be significant when $z_p<z_p^{(1)}$. 
In addition, as the positive links become too dense meaning that $z_p$ is larger than
the second threshold $z_p^{(2)}$, i.e., $z_p^{(2)} \simeq4.3$ for $z_n=0$, nodes that exceed 
their threshold become too rare, so that the global cascade can hardly occur \cite{watts2002}. 
As a result, the size $\rho$ of global cascades drops 
abruptly to zero regardless of $z_n$ for large $z_p \gtrsim z_p^{(2)}$. 
When the value of average degree $z_p$ is in between, $z_p^{(1)} < z_p < z_p^{(2)}$, the global 
cascade can occur. In this regime, negative links play an important role in preventing 
the activation of nodes connected to active nodes, so that the size of global cascades 
$\rho$ noticeably decreases as $z_n$ increases. 
Note that the second threshold $z_p^{(2)}$ is $\theta$-dependent.

Figure~\ref{fig:rho}(c) shows the decrease of $\rho$ as $z_n$ increases for various 
$z_p=1$ ($\circ$),  $z_p=2$ ($\triangle$), $z_n=3$ ($\triangledown$), $z_n=4$ ($\diamond$), 
$z_p=5$ ($\pentagon$). We again confirm the monotonic decrease of $\rho$ with 
an increasing number of negative links. To quantitatively measure this 
effect, we measured the difference between $\rho$ at $z_n$ and $\rho$ in the absence 
of negative links, denoted by $|\Delta \rho|$. Specifically, the gap $|\Delta \rho|$ 
is defined to be $|\Delta \rho(z_n)|= |\rho(z_n)-\rho(z_n=0)|$. The gap $|\Delta \rho|$ 
represents how much negative links depress cascading activations compared to
the absence of negative links. The inset of Fig.~\ref{fig:rho}(c) shows $|\Delta \rho|$ for
$z_p=2$ ($\triangle$), $z_n=3$ ($\triangledown$), $z_n=4$ ($\diamond$) with
increasing $z_n$ in double logarithmic scale. While $z_n$ is less than unity, 
$|\Delta \rho|$ increase linearly with increasing $z_n$ such that 
$|\Delta \rho| \sim z_n$. Above $z_n\gtrsim 1$, networks contain loops formed by 
negative links, so that the increase of $\Delta \rho$ slows down and 
becomes no longer linear with respect to $z_n$.

In addition to the final size $\rho$ of global cascades, we also measured the duration $T$ of 
cascading dynamics [Fig.~\ref{fig:rho}(d-f)]. When the cascades of activation start 
from seed nodes, the cascade size increases rapidly in the early time and finally 
reaches $\rho$ in the steady state. The duration of the global cascade, $T$, is the time 
from when the cascading first starts to when it falls into a steady state. We computed the duration 
$T$ for different sets of $z_n$ and $z_p$. We found that the peaks of the duration 
$T$ are located near the transition points where global cascades emerge or disappear.
A similar phenomenon which is related to the critical slowing down was 
observed in the previous studies of threshold models \cite{kmlee2014,parshani2011}.

\subsection{Theory: mean-field approximation}

We next develop a mean-field-type analytic approximation in order to estimate $\rho$ for given
degree distribution and seed fraction. When there is no negative link, the 
global cascade size can be analytically calculated by using the generating function 
method~\cite{watts2002,gleeson2007,gleeson_porter_book}. For signed networks, however, the effect of 
negative links must be included. It is important to note that the aforementioned 
`blocking' effect is applied only in one direction and the directionality is 
determined dynamically according to the order in which the state of nodes is 
determined. Such a direction cannot be exactly determined based on structural 
information alone. Therefore we apply the following simple approximation. 
In random networks, a zero-th order estimation for the probability that one node in a pair connected by a 
negative link will be determined for its state first is uniform as $1/2$. 
Then, we can assume that there are $k_n/2$ negative links that effectively 
act on a node having $k_n$ number of negative neighbors.

Using the above approximation, we derived the self-consistency equations
for the probability $q_p$ ($q_n$) that a node reached along a randomly 
chosen positive (negative) link is active. The probabilities $q_p$ and $q_n$ 
satisfy 
\begin{align}
q_p =&\,\rho_0 + (1-\rho_0  - \rho_0 z_n)\sum_{k_p, k_n} \frac{k_p P(k_p, k_n)}{\langle k_p \rangle}  \\
&\times \sum_{r=0}^{k_p-1} \binom{k_p-1}{r} q_p^{r}(1-q_p)^{k_p-1-r} F\left(\frac{r}{k_p} \right) (1-q_n)^{k_n/2}, \nonumber \\
q_n =&\,\rho_0+ (1-\rho_0  - \rho_0 z_n)\sum_{k_p, k_n} \frac{k_n P(k_p, k_n)}{\langle k_n \rangle}  \\
&\times \sum_{r=0}^{k_p} \binom{k_p}{r} q_p^{r}(1-q_p)^{k_p-r}  F\left(\frac{r}{k_p} \right) (1-q_n)^{(k_n-1)/2}, \nonumber 
\end{align}
where $P(k_p,k_n)$ is the degree distribution of the network and $F(r/k_p)$ is the 
activation function. $F(r/k_p)$ is $1$ for $r/k_p>\theta$ where $r$ is the number 
of active positive neighbors and $F(x)=0$ otherwise. The term $\rho_0 z_n$ 
approximately corresponds to the probability that the fraction of nodes that 
are blocked by seed nodes, for a small $\rho_0$. 
Once obtaining $q_p$ and $q_n$ by simultaneously solving Eqs.~(1) and (2), we can estimate the  size $\rho$ of global cascades as
\begin{align}
\rho =&\,\rho_0  +(1-\rho_0  - \rho_0 z_n)\sum_{k_p, k_n} P(k_p, k_n) \\
 &\times  \sum_{r=0}^{k_p} \binom{k_p}{r} q_p^{r}(1-q_p)^{k_p-r} F\left(\frac{r}{k_p} \right) (1-q_n)^{k_n/2}. \nonumber
\end{align}

\begin{figure}[t]
\includegraphics[width=0.99\linewidth]{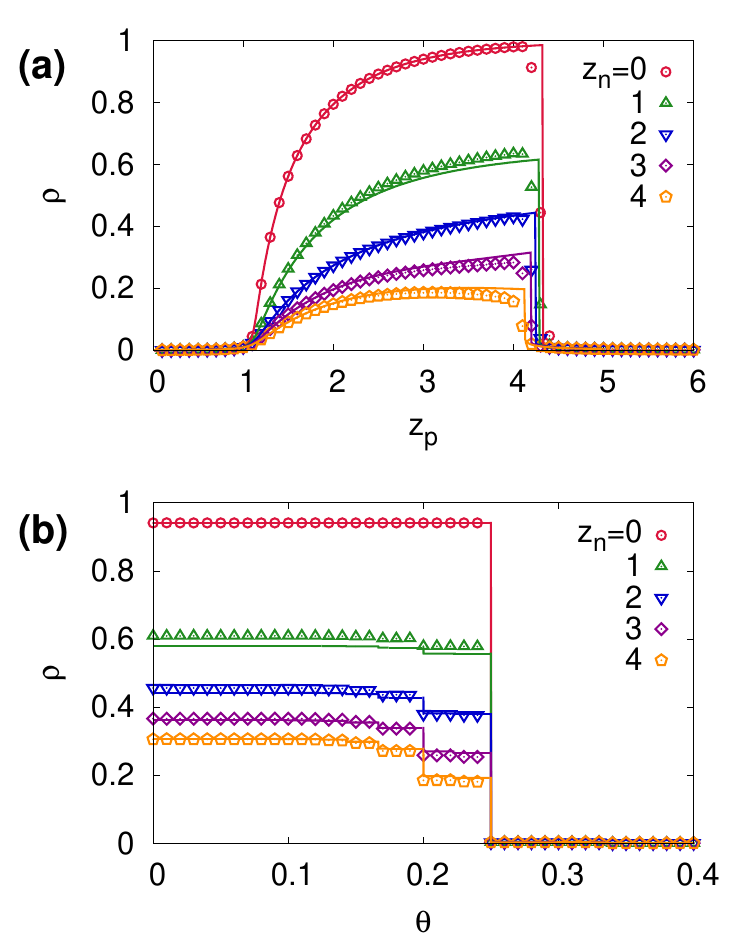}
\caption{
Plots of the mean-field-theory-obtained cascade size $\rho$ (solid lines) on signed ER networks 
as a function of (a) positive-link mean degree $z_p$ and $\theta=0.2$ and (b) threshold $\theta$ assigned to each 
individual for $z_p=3$, with various values of negative-link mean degree $z_n=0,1,2,3,4$.
The theoretical results show good agreement with numerical simulation results shown in symbols.
\label{fig:th}
}
\end{figure}

We found that the size of global cascades predicted by the theory shows good 
agreement with the numerical results as shown in Fig.~\ref{fig:th}.
Figure \ref{fig:th}(a) shows the size $\rho$ of global cascades as a function of $z_p$ with
various $z_n=0$ ($\circ$),  $z_n=1$ ($\triangle$), $z_n=2$ ($\triangledown$), $z_n=3$ 
($\diamond$), $z_n=4$ ($\pentagon$).
The theory implies that the condition for global cascades becomes \cite{watts2002,gleeson2007}
\begin{align}
\sum_{k_p=1}^\infty \frac{k_p(k_p-1)}{z_p} P(k_p) F\left(\frac{1}{k_p}\right) > \frac{1}{1-\rho_0-\rho_0 z_n},
\label{eq:condition}
\end{align}
where $F(1/k)$ is 1 if $1/k>\theta$ and 0 otherwise.
Therefore the theory predicts that the first transition $z_p^{(1)}$ from no cascading 
to global cascades is located near $z_p^{(1)} =1/(1-\rho_0-\rho_0 z_n)$, which is less 
sensitive to $z_n$ for a small $\rho_0$. 
In addition, the second threshold $z_p^{(2)}$ from the global cascade to no cascade phases 
decreases with increasing $z_n$ because the number of nodes that are initially inactive 
increases with increasing $z_n$ by the term $\rho_0 z_n$.  
Figure \ref{fig:th}(b) shows the behaviour of $\rho$ as a function of the threshold $\theta$ 
with $z_p=3$ and various $z_n=0$ 
($\circ$),  $z_n=1$ ($\triangle$), $z_n=2$ ($\triangledown$), $z_n=3$ ($\diamond$), 
$z_n=4$ ($\pentagon$). The theory successfully predicts the cascade condition for the threshold 
from Eq.~(\ref{eq:condition}) and the suppression of the global cascades with increasing $z_n$.

\subsection{Heterogeneity in Cascades of Activations}

\begin{figure}[t]
\centering
\includegraphics[width=\linewidth]{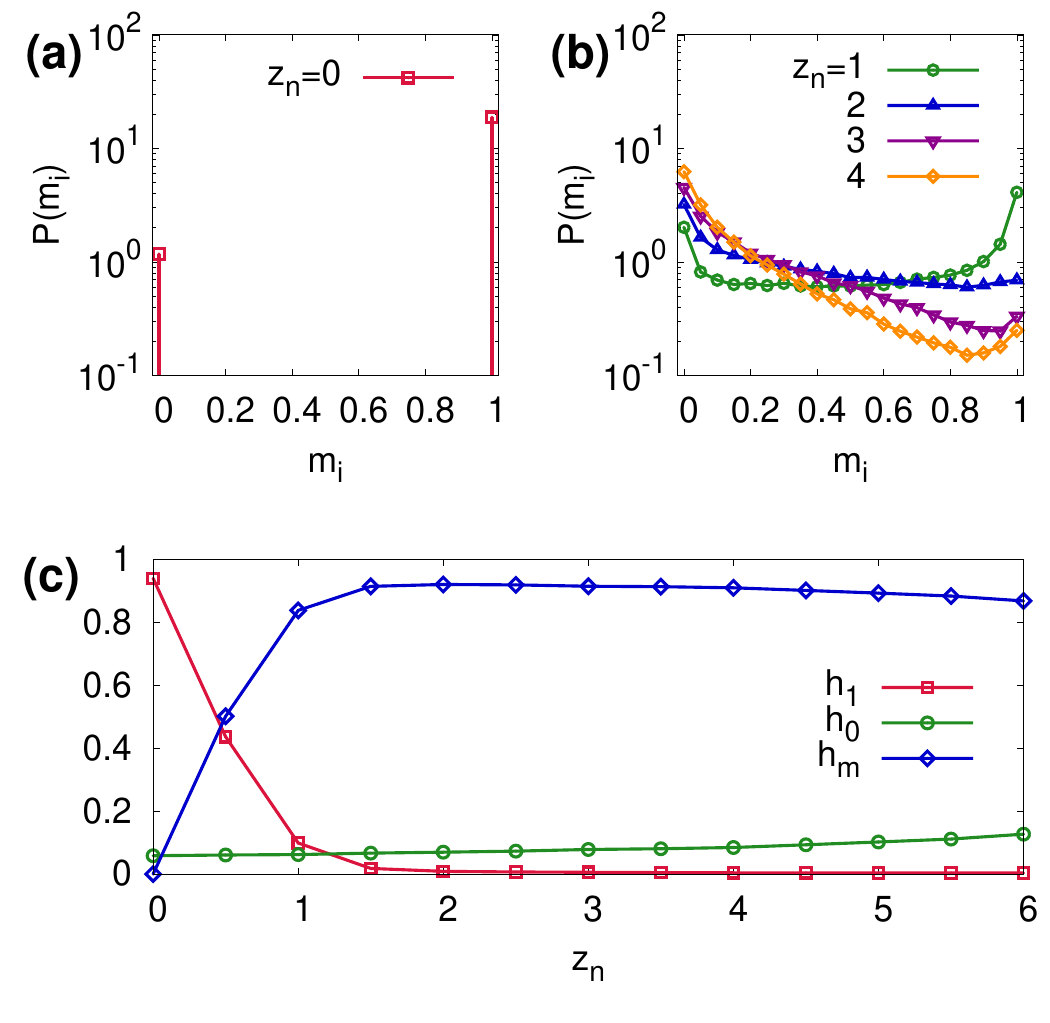}
\caption{
(a,b) The probability distribution of activation frequency $P(m_i)$ for 
(a) $z_{n}=0$ and (b) $z_n=1,2,3,4$ on ER networks with $N=10^5$ and $z_{p}=3.0$. 
(c) The fraction of nodes that are always active $h_1$, never active $h_0$, and 
occasionally active $h_m$ as a function of $z_n$, for an ER network with $N=10^5$, 
$\rho_0=10^{-3}$, and $z_p=3$, averaged over $10^4$ realizations with different 
sets of initial seeds. 
}
\label{fig:hetero}
\end{figure}

Negative links in signed networks not only suppress global cascades
but also render the patterns of activations more heterogeneous. When there is no 
negative link ($z_n=0$), the final configuration of activations is uniquely defined 
for a given set of seeds on a given network, meaning 
that nodes are either always activated or never activated in different trials of cascading dynamics.
When $z_n>0$, however, the final configuration resulting from the cascade dynamics is not uniquely determined 
even though the dynamics starts from the same network structure as exemplified in Fig.~\ref{fig:model}. 
In other words, each node in the signed networks responds differently to the dynamics depending on the sequence 
of the cascades of activations. To quantify such heterogeneity, we define $m_i$ as the probability 
that node $i$ is active over different realizations on the same network configuration. 
We measured the probability distribution of $m_i$ for a given network structure. 
When $z_n=0$, $P(m_i)$ is simply given by the two peaks at $m_i=0$ and $m_i=1$ 
[see Fig. \ref{fig:hetero}(a)]. However, when $z_{n}>0$ most nodes are active 
only occasionally meaning that $0<m_i<1$. 
As shown in Fig.~\ref{fig:hetero}(b), sporadically active nodes are widely 
distributed in $m_i$, accompanied by the two peaks at $m_i=0$ and $1$. 
In addition, the latter peak at $m_i=1$ is substantially weaker than that of $z_n=0$ case. 

The response of nodes to cascading dynamics can fall into three classes: 
always active ($m_i=1$), never active ($m_i=0$), and occasionally active ($0<m_i<1$). 
We measured the fraction of nodes that are always active $h_1$, never active $h_0$, and 
occasionally active $h_m$ as a function of $z_n$, for 
a fixed signed ER network of 
$z_p=3$ with an ensemble of random initial seed sets [Fig.~\ref{fig:hetero}(c)]. 
The sum of the three fractions satisfies the sum rule $h_0 + h_m + h_1 =1$. With increasing $z_n$, 
the fraction $h_1$ of the always active nodes  decreases rapidly and becomes 
almost negligible when $z_n \gtrsim 2$. 
When the density of negative links increases, most nodes in a network belong to 
occasionally active state $h_m$ because of the negative links. 
The fraction $h_0$ of nodes that are never activated maintains a low fraction 
for all range of $z_n$, but increases steadily as $z_n$ increases.
The observation that the $h_m$-state nodes constitute a dominant fraction indicates 
the prevalence of system-wide heterogeneity of dynamics driven by the negative links.

Finally, in order to assess quantitatively the heterogeneity for different realizations of 
activation patterns, we computed the average overlap 
between different `samples' (activation configurations) defined by \cite{parisibook,folena}
\begin{align}
Q = \left\langle \sum_{i=1}^N s_i^{\alpha} s_i^{\beta} \right\rangle_{\alpha\beta}, 
\end{align}
where $\alpha$ and $\beta$ represent the indices of samples. 
Here a sample refers to the particular outcome of signed cascade dynamics on the same 
network structure but with random initial seed sets. 
$\langle \cdot \rangle_{\alpha\beta}$ represents the sample average.
$s_i^{\alpha}$ represents the state of node $i$ at the steady state in sample $\alpha$, i.e, 
$s_i^{\alpha}=1$ and $0$ respectively represents that node $i$ is active and inactive in sample $\alpha$. 
(Note that in the spin-glass literature \cite{parisibook,folena} the same formula for the overlap is 
used but with the Ising spin variables $s_i=\pm1$.)

\begin{figure}[t]
\centering
\includegraphics[width=\linewidth]{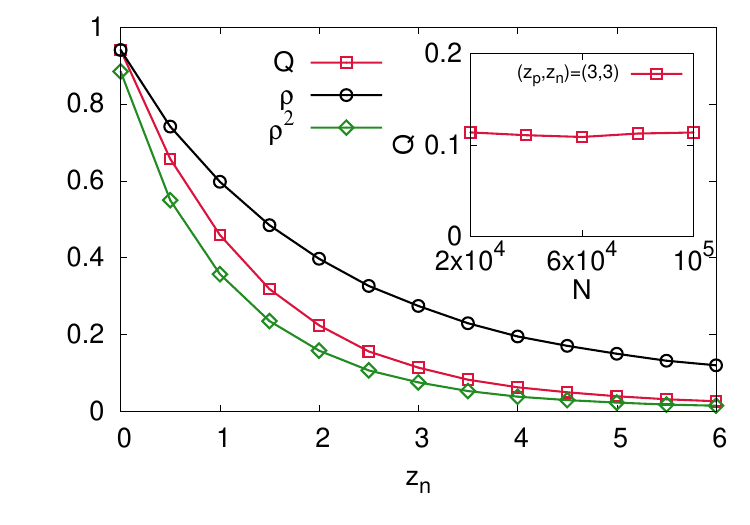}
\caption{
The overlap $Q$ averaged over $10^3$ different realizations of activation patterns 
for ER networks with $N=10^5$, $\rho_0=10^{-3}$, and $z_p=3$. For comparison, $\rho$ 
and $\rho^2$ are also shown. Inset shows the overlap $Q$ with $(z_p,z_n)=(3,3)$ 
for different network size $N$ from $2\times 10^4$ to $10^5$.
}
\label{fig:overlap}
\end{figure}

The overlap would become maximized as $Q=\rho$ when the activation patterns for 
each node are completely identical for different realizations. On the other hand, if the activation patterns 
are completely uncorrelated for different realizations, the overlap $Q$ would become $Q=\rho^2$.
As shown in Fig.~\ref{fig:overlap}, $Q$ is maximized at $z_n=0$ exhibiting $Q \approx \rho$
and decreases steadily as $z_n$ increases. When $z_n>0$, $Q$ is less than $\rho$ 
but greater than $\rho^2$, meaning that the configurations are correlated, but partially. 
Moreover, as $z_n$ increases, the different activation patterns become more and more 
uncorrelated. This can be inferred as the result that $Q$ approaches to $\rho^2$ with increasing $z_n$. 
In addition, we can see in the inset of Fig.~\ref{fig:overlap} that the overlap $Q$ remains a constant value regardless 
of the increase of system size $N$. The heterogeneity of the activation patterns suggests that 
many different activation patterns can come from the same network structure. This is reminiscent 
of infinitely many ground states in disordered spin systems \cite{parisibook}. 
It also implies that the final fate of each node cannot be fully predicted from the structure of networks.

\subsection{On signed scale-free networks}

\begin{figure}[t]
\centering
\includegraphics[width=\linewidth]{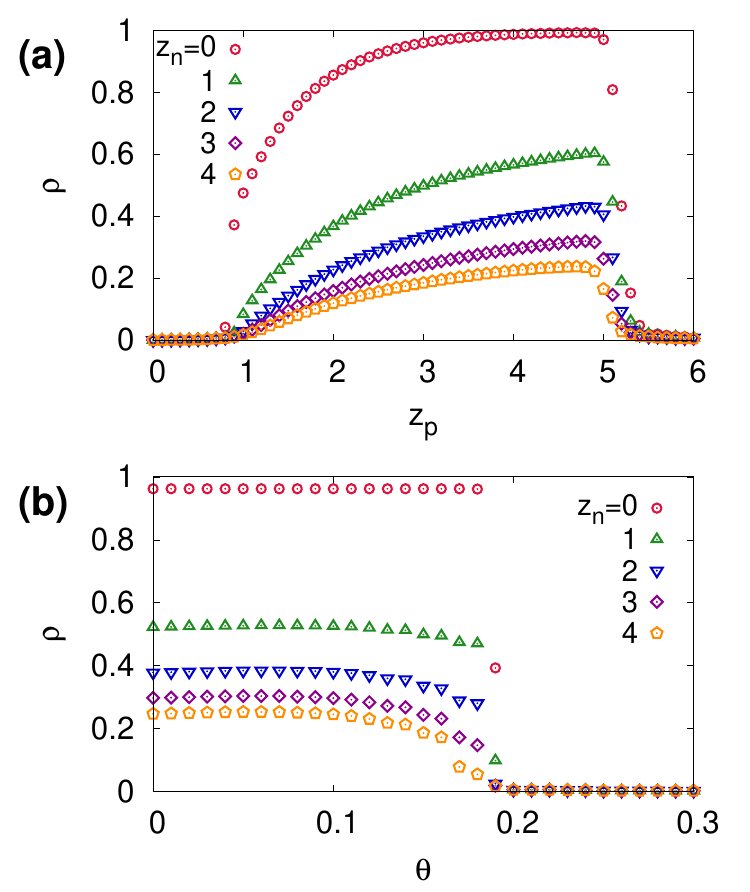}
\caption{
The cascade size $\rho$ on signed SF networks with $N=10^5$, degree exponent $\gamma=2.5$,
seed fraction $\rho_0=10^{-3}$, averaged over $10^2$ different network ensembles
as a function of (a) positive-link mean degree $z_p$ with $\theta=0.15$ and (b) threshold $\theta$
with $z_p=3$. 
}
\label{fig:sfnet}
\end{figure}

In order to check the effect of negative links on signed heterogeneous networks,
we study the signed cascade model on signed scale-free (SF) networks.
For building signed SF networks, we use static SF network model \cite{static}. 
Initially, there are $N$ isolated nodes, i.e., $N=10^5$ in our study.
By following rules, we constructed signed static SF networks.
Each node $i$ has its inherent weight $\omega_i$ given by 
$\omega_i = i^{-\mu} / \sum_{j=1}^N j^{-\mu}$,
where $\mu$ is a constant, $0 < \mu < 1$, which determines the degree exponent.
We choose a pair of nodes, say $i$ and $j$ independently following the probability $\omega_i$ 
and $\omega_j$, and connect them unless they are already connected, for both 
positive and negative links. 
We repeat the procedures until the mean degree reaches $z_p$
for positive links and $z_n$ for negative links.
The degree distribution of resulting networks is asymptotically scale-free with 
the tail decaying as $\sim(k_p+k_n)^{-\gamma}$ with the degree exponent $\gamma = (\mu+1)/\mu$.
Note that the degree distribution for positive or negative links respectively also has the same power-law tail and the positive degree $k_p$ and negative degree $k_n$ of a given node are correlated in this signed static SF networks.

We perform numerical simulations of signed cascade model on the signed static SF networks
with the degree exponent $\gamma =2.5$. 
Figure \ref{fig:sfnet}(a) shows the size $\rho$ of global cascades as a function of $z_p$ with
various $z_n=0$ ($\circ$),  $z_n=1$ ($\triangle$), $z_n=2$ ($\triangledown$), $z_n=3$ 
($\diamond$), $z_n=4$ ($\pentagon$) with $\theta=0.15$. 
Global cascade can occur under the condition where $z_p^{(1)} < z_p < z_p^{(2)}$
as the signed ER networks. In this regime, we confirm the suppression of the global 
cascades with increasing $z_n$ in signed scale-free networks. 
Figure \ref{fig:th}(b) shows the behaviour of $\rho$ as a function of threshold $\theta$ with $z_p=3$ and various $z_n=0$ ($\circ$),  $z_n=1$ ($\triangle$), $z_n=2$ ($\triangledown$), $z_n=3$ 
($\diamond$), $z_n=4$ ($\pentagon$). 
We found that $\rho$ abruptly decreases at the critical threshold.
Comparing with the cases on signed ER networks [Fig.~3(b)], the global cascade disappears at a lower value of $\theta$.  
To summarize, we observe qualitatively similar features on signed SF networks to those on signed ER networks: we confirm that the negative links suppress the size $\rho$ of global cascade in scale-free networks while the critical threshold is insensitive to $z_n$.

\subsection{On signed real-world networks}

We test the effect of negative links on signed ``real-world'' networks from the data of online social media.
By using the real-world signed network data, we can not only study the model on more realistic setting,
but also address the effect of higher-order structural features such as structural  balance
and degree correlations.
We constructed signed networks from the {\it Epinions} and {\it Slashdot} datasets obtained from Stanford
Network Analysis Project (SNAP)~\cite{leskovec2010, snap_url}. {\it Epinions} is an online review web site 
where consumers can have signed relationships one another  based on trust or distrust. {\it Slashdot} is a 
technology-related news blog where users can tag each other as friends and foes. 
When creating signed networks, we converted directed links in the raw data into 
undirected links, so that the networks become undirected.
The Epinions network has $131,828$ nodes with $z_p=8.96$ and $z_n=1.84$,
and the slashdot network has $82,140$ nodes with $z_p=9.27$ and $z_n=2.91$.

\begin{figure}[t]
\centering
\includegraphics[width=1.0\linewidth]{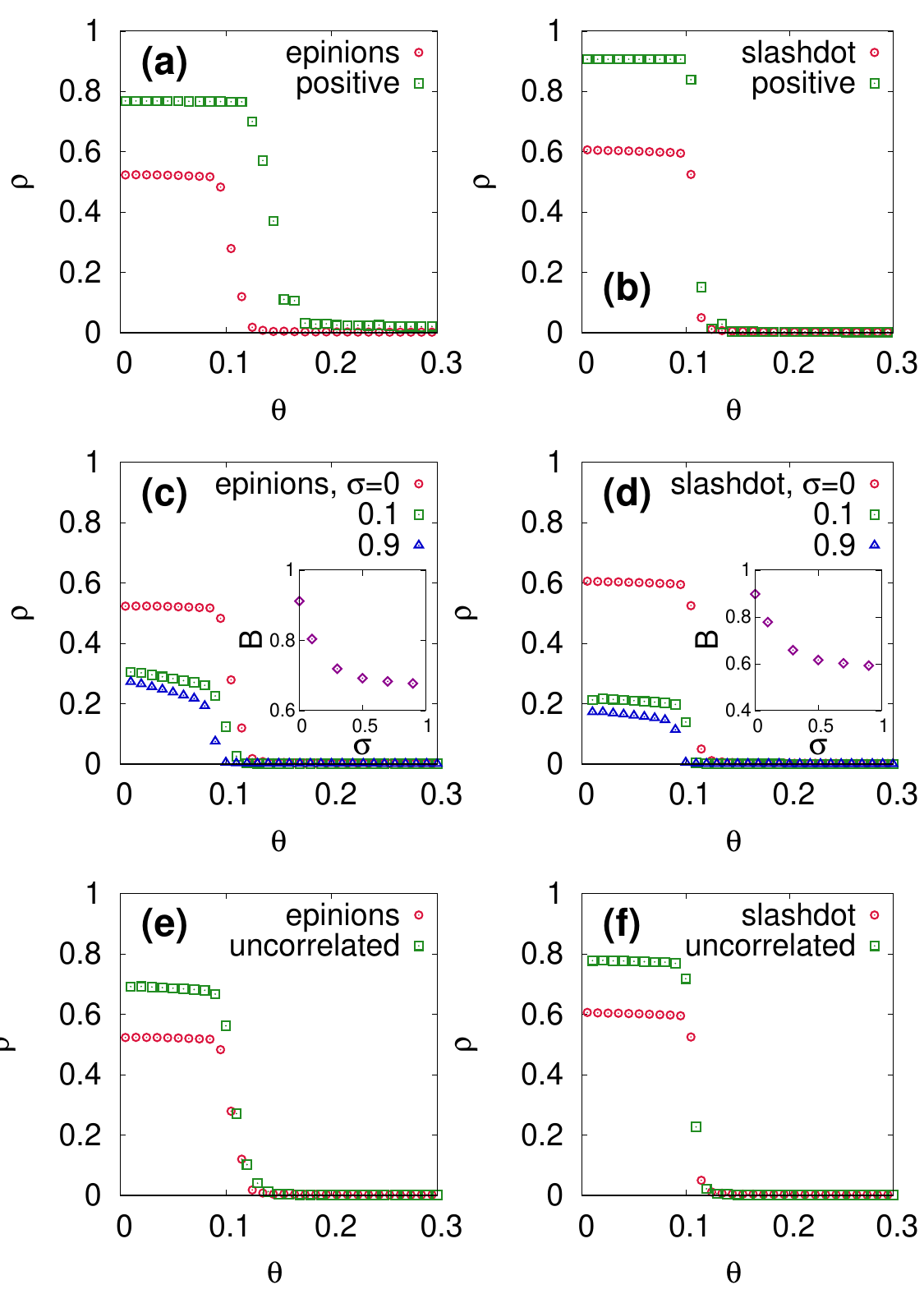}
\caption{
The size $\rho$ of global cascades on signed real-world networks, (a) {\it epinions} and 
(b) {\it slashdot} networks, as a function of threshold $\theta$. 
We plot together the size $\rho$ of cascades in real-world networks with solely positive links for comparison. 
Plots of the cascade size $\rho$ of the sign shuffled surrogates preserving network structures 
of (c) {\it epinions} and (d) {\it slashdot} networks as a function of threshold $\theta$.
Inset shows the fraction of balance triads as a function of the ratio $\sigma$ of shuffled links. 
Plots of the cascade size $\rho$ of the degree shuffled surrogates that remove 
the degree correlation between positive and negative links of (e) {\it epinions} 
and (f) {\it slashdot} networks as a function of threshold $\theta$.
}
\label{fig:real}
\end{figure}

First, we perform numerical simulations of signed cascade processes in the two real-world networks
by varying the threshold $\theta$. Figures \ref{fig:real}(a,b) show the size $\rho$ of 
global cascades as a function of $\theta$ in {\it Epinions} and {\it Slashdot} networks. 
To investigate the effect of negative links, we also calculated the size $\rho$ of global 
cascades in real-world networks with only positive links after removing all negative links.
We confirm that negative links still significantly reduce the global cascade size 
in empirical networks by comparing $\rho$ when the negative links are removed.

Second, we explore the effect of the structural balance using the real-world signed networks.
In order to generate the signed network structures with different levels of structural balance, we take the empirical signed network data and perform the ``sign-shuffling'' while preserving the overall network connections intact to the original data. To be specific, we select a pair of links (irrespectively of signs) in the network uniformly at random and exchange their signs. 
The insets of Figs.~\ref{fig:real}(c,d) show that the fraction $B$ of balanced triads, a measure of the structural balance level, decreases monotonically with the number of edge-shufflings per link $\sigma$.  
We measure the size $\rho$ of global cascade as we perform the edge-shuffling. 
Figures \ref{fig:real}(c,d) compare the case of $\sigma = 0$ (original unshuffled network), $\sigma=0.1$, and $\sigma =0.9$ for the real-world signed networks, showing that the size of global cascade decreases with $\sigma$.

Finally, we measure the size $\rho$ of the global cascade in ``randomly-coupled'' signed surrogates of 
empirical networks to evaluate the impact of the correlation between positive-link and negative-link degrees of a given node.
We created the randomized surrogates that removes the degree correlation between the positive and 
negative links in the original network.
Figures \ref{fig:real}(e,f) shows $\rho$ of the original network
and randomized networks as a function of $\theta$ for {\it Epinions} and {\it Slashdot} networks. 
The size of global cascade is greater in randomized networks without degree correlation than in original network as shown in Fig.~\ref{fig:real}(e,f). 

\section{Conclusion}

To conclude, in this paper we proposed and studied a threshold cascade model on signed networks with both positive 
and negative interactions. In our model, we impose the role of negative links by the rule that a node's activation 
is completely blocked if there is at least one active neighbor connected by a negative link. 
We found that the negative links not only suppress the global cascades but also produce the heterogeneity 
in the activation patterns. As the number of negative links increases, the activation patterns of threshold model
become increasingly uncorrelated. The mean-field-type approximation theory developed for the effects of negative 
links successfully accounts for the observed results. 
We also address the impact of network 
structural balance and degree correlation between positive and negative links to the global cascade size on 
a signed network using real-world networks and their randomized surrogates.
Our results imply that the inhibitory interactions which 
are widespread in many real-world systems may play not only to suppress the activations but also to render 
variability in dynamical patterns.

The model studied in this study is a toy model illustrating the essential functional role of negative 
links in threshold cascade dynamics, offering many rooms for extensions over its limitations. 
For example, from the model perspective, a rather relaxed or probabilistic form of suppressive function 
of negative links can be considered; a mixture of the currently-implemented multiplicative suppression 
and the additive suppression as considered in Ref.~\cite{li2021} is also worth the investigation.
From signed-network structural perspective, more comprehensive study on the effects of higher-order 
organization of signed interactions such as social balance at the triad level and beyond needs 
systematic examination, which is by no means a trivial problem. From theoretical perspective, further 
understanding of similarities to and differences from spin-glass phenomenology might be of interest.
As a final outlook, we anticipate the functional relevance and dynamic importance of negative 
links showcased in this paper to apply not only to the simple social dynamic processes considered 
in this study but also to many other different dynamical processes on signed networks, including 
brain, offering an ample arena for future works.

\section{Acknowledgments}
This work was supported in part by the National Research Foundation of Korea (NRF) grants funded by the Korea 
government (MSIT) (No.~2020R1A2C2003669 (K-IG) and No.~2020R1I1A3068803 (BM)). 
SL was supported by Basic Science Research Program through the National Research Foundation of 
Korea (NRF) funded by the Ministry of Education (No.~2016R1A6A3A11932833). 
K-IG would also like to thank the APCTP for its hospitality during the completion of this work. \\


\begin{thebibliography}{99}
\bibitem{schelling} T. Schelling, Hockey helmets, concealed weapons, and daylight saving: A study of binary choices with externalities, J. Conflict Resolution {\bf 17}, 381-428 (1973).
\bibitem{granovetter} M. Granovetter, Threshold models of collective behavior, Am. J. Soc. {\bf 83}, 1420-1443 (1978).
\bibitem{watts2002} D.~J.~Watts, A simple model of global cascades on random networks, Proc.~Natl.~Acad.~Sci.~USA {\bf 99}, 5766 (2002).
\bibitem{rohlf2002} T. Rohlf and S. Bornholdt, Criticality in random threshold networks: annealed approximation and beyond, 
	Physica A, {\bf 310}, 245-259 (2002).
\bibitem{friedman2012} N. Friedman, S. Ito, B. A. W. Brinkman, M. Shimono, R. E. Lee DeVille, K. A. Dahmen, J. M. Beggs, and T. C. Butler, Universal Critical Dynamics in High Resolution Neuronal Avalanche Data, Phys. Rev. Lett. {\bf 108}, 208102 (2012).
\bibitem{kusmierz2020} \L{}. Ku\'smierz, S. Ogawa, and T. Toyoizumi, Edge of Chaos and Avalanches in Neural Networks with Heavy-Tailed Synaptic Weight Distribution, Phys. Rev. Lett. {\bf 125}, 028101 (2020).
\bibitem{kmlee2011} K.-M. Lee, J.-S. Yang, G. Kim, J. Lee, K.-I. Goh, and I.-m. Kim, Impact of the topology of global macroeconomic network on the spreading of economic crises, PloS one {\bf 6}, e18443 (2011).
\bibitem{buldyrev2010} S. V. Buldyrev, R. Parshani, G. Paul, H. E. Stanley, and S. Havlin, Catastrophic cascade of failures in interdependent networks, Nature {\bf 464} 1025-1028 (2010).
\bibitem{brummitt2012a} C. D. Brummitt, R. M. D'Souza, and E. A. Leicht, Suppressing cascades of load in interdependent networks, Proc. Natl. Acad. Sci. USA {\bf 109}, E680-E689 (2012).
\bibitem{bmin2014} B. Min and K.-I. Goh, Multiple resource demands and viability in multiplex networks, Phys. Rev. E {\bf 89}, 040802(R) (2014).
\bibitem{gleeson2007} J. P. Gleeson and D. J. Cahalane, Seed size strongly affects cascades on random networks, Phys. Rev. E {\bf 75}, 056103 (2007).
\bibitem{motter2002} A. E. Motter and Y. C. Lai, Cascade-based attacks on complex networks, Phys. Rev. E {\bf 66}, 065102 (2002).
\bibitem{hackett2011} A. Hackett, S. Melnik, and J. P. Gleeson, Cascades on a class of clustered random networks, Phys. Rev. E {\bf 83}, 056107 (2011).
\bibitem{brummitt2012b} C. D. Brummitt, K.-M. Lee, and K.-I. Goh, Multiplexity-facilitated cascades in networks, Phys. Rev. E {\bf 85}, 045102(R) (2012).
\bibitem{kmlee2014} K.-M. Lee, C. D. Brummitt, and K.-I. Goh, Threshold cascades with response heterogeneity in multiplex networks, Phys. Rev. E {\bf 90}, 062816 (2014).
\bibitem{bowers2004} P. M. Bowers, S. J. Cokus, D. Eisenberg, and T. O. Yeates, Use of logic relationships to decipher protein network organization, Science {\bf 306}, 2246-2249 (2004).
\bibitem{szell2010} M. Szell, R. Lambiotte, and S. Thurner, Multirelational organization of large-scale social networks in an online world. Proc. Natl. Acad. Sci. USA {\bf 107}, 13636 (2010).
\bibitem{leskovec2010a} J. Leskovec, D. Huttenlocher, and J. Kleinberg, Signed networks in social media, Proc. of the SIGCHI Conf. on Human Factors in Computing Systems, 1361 (2010).
\bibitem{rubinov2010} M. Rubinov and O. Sporns, Complex network measures of brain connectivity: Uses and interpretations, Neuroimage {\bf 52}, 1059 (2010).
\bibitem{tang2016} J. Tang, Y. Chang, C. Aggarwal, and H. Liu, A survey of signed network mining in social media, ACM Computing surveys {\bf 49} 1-37, (2016).
\bibitem{leskovec2010b} J. Leskovec, D. Huttenocher, and J. Kleinberg, Predicting positive and negative links in online social networks, Proc. of the 19th international conference on WWW. ACM (2010).
\bibitem{bjkim} H. J. Park, S. D. Yi, D. J. Kim, and B. J. Kim, Network of likes and dislikes: Conflict and membership, Physica A {\bf 461}, 647 (2016). 
\bibitem{regulondb} S. Gama-Castro et. al, RegulonDB version 9.0: high-level integration of gene regulation, coexpression, motif clustering and beyond, Nucleic Acids Res. {\bf 44}, D133 (2015).
\bibitem{regulondbv10} S. Santos-Zavaleta et. al, RegulonDB v 10.5: tackling challenges to unify classic and high throughput knowledge of gene regulation in E. coli K-12, Nucleic Acids Res. {\bf 47}, D212 (2019).
\bibitem{maozbook} Z. Maoz, (2010) {\it Networks of nations: The evolution, structure, and impact of international networks, 1816-2001}, Vol. 32. Cambridge Univ. Press.
\bibitem{maoz2007} Z. Maoz, L. G. Terris, R. D. Kuperman, and I. Talmud, What is the enemy of my enemy? Causes and consequences of imbalanced international relations, 1816-2001, Journal of Politics {\bf 69}, 100 (2007).
\bibitem{parisibook} M. M\'ezard, G. Parisi, and M. Virasoro, {\it Spin Glass Theory And Beyond} (World Scientific, Singapore, 1986). 
\bibitem{heider1946} F. Heider, Attitudes and cognitive organization, The Journal of Psychology {\bf 21}, 107 (1946).
\bibitem{cartwright1956} D. Cartwright and F. Harary, Structural balance: a generalization of Heider's theory, Psychological review {\bf 63}, 277 (1956).
\bibitem{facchetti2011} G. Facchetti, G. Iacono, and C. Altafini, Computing global structural balance in large-scale signed social networks, Proc. Natl. Acad. Sci. USA {\bf 108} 20953 (2011).
\bibitem{he2018} X. He, H. Du, M. Cai, M. W. Feldman, The evolution of cooperation in signed networks 
	under the impact of structural balance, Plos One, 13(10):e0205084, (2018).	
\bibitem{du2016} H. Du, X. He, M. W. Feldman, Structural balance in fully signed networks, Complexity 21(S1): 497–511 (2016).
\bibitem{ciotti2015} V. Ciotti, G. Bianconi, A. Capocci, F. Colaiori, and P. Panzarasa, Degree correlations in signed social networks, Physica A (Amsterdam) {\bf 422}, 25 (2015).
\bibitem{dhkim2005} D.-H. Kim, G. J. Rodgers, B. Kahng, and D. Kim, Spin glass transitions on scale-free networks, Phys. Rev. E {\bf 71}, 056115 (2005). 
\bibitem{antal2005} T. Antal, P. Krapivsky, and S. Redner, Dynamics of social balance on networks, Phys. Rev. E {\bf 72}, 036121 (2005).
\bibitem{antal2006} T. Antal, P. Krapivsky, and S. Redner, Social balance on networks: The dynamics of friendship and enmity, Physica D {\bf 224}, 130 (2006).
\bibitem{zhao2013a} K. Zhao, and G. Bianconi, Percolation on interacting, antagonistic networks, J. Stat. Mech. {\bf 2013}, P05005 (2013).
\bibitem{quattrociocchi2014} W. Quattrociocchi, G. Caldarelli, and A. Scala, Opinion dynamics on interacting networks: media competition and social influence, Sci. Rep. {\bf 4}, 4938 (2014). 
\bibitem{saeedian2016} M. Saeedian, N. Azimi-Tafreshi, G. R. Jafari, and J. Kertesz, Epidemic spreading on evolving signed networks, Phys. Rev. E {\bf 95}, 022314 (2017).
\bibitem{li2021} H.-J. Li, W. Xu, S. Song, W.-X. Wang, and M. Perc, The dynamics of epidemic spreading on signed networks, Chaos, Solitons, \& Fractals, {\bf 151}, 111294 (2021).
\bibitem{nowak2019} B. Nowak and K. Sznajd-Weron, Homogeneous symmetrical threshold model with nonconformity: Independence versus anticonformity, Complexity {\bf 2019}, 1 (2019)
\bibitem{nowak2022} B. Nowak, M. Grabisch, and K. Sznajd-Weron, The threshold model with anticonformity under random sequential updating, Phys. Rev. E {\bf 105}, 054314 (2022).
\bibitem{he2019} X. He, H. Du, M. W. Feldman, and G. Li, Information diffusion in signed networks, Plos One, 14(10):e0224177 (2019).
\bibitem{pozveh2019} M. Hosseini-Pozveh, et. al., Assessing information diffusion models for 
	influence maximization in signed social networks, Expert Systems with Applications 119, 476 (2019).
\bibitem{li2013} Y. Li, et. al., Influence diffusion dynamics and influence maximization in social networks with friend 
	and foe relationships, Proceedings of the sixth ACM international conference on Web search and data mining. (2013).
\bibitem{li2019} D. Li and L. Jiming, Modeling influence diffusion over signed social networks, 
	IEEE Transactions on Knowledge and Data Engineering {\bf 33(2)}, 613-625 (2019).
\bibitem{qu2021} C. Qu, J. Bi, and G. Wang. Personalized information diffusion in signed social networks, 
	Journal of Physics: Complexity {\bf 2(2)}, 025002 (2021). 
\bibitem{molloy1998} M. Molloy and B. Reed, The size of the giant component of a random graph with a given degree sequence, Combinatorics, Probability and Computing {\bf 7}, 295-305 (1998).
\bibitem{parshani2011} R. Parshani, S. V. Buldyrev, and S. Havlin, Critical effect of dependency groups on the function of networks, Proc.~Natl.~Acad.~Sci.~USA {\bf 108(3)}, 1007-1010 (2011).
\bibitem{gleeson_porter_book} M. A. Porter and J. P. Gleeson, {\it Dynamical Systems on Networks} (Springer, Heidelberg, 2016). 
\bibitem{static} K.-I. Goh, B. Kahng, and D. Kim, Universal behavior of load distribution in scale-free networks,
	Phys. Rev. Lett. {\bf 87}, 278701 (2001).
\bibitem{leskovec2010} J. Leskovec, D. Huttenlocher, and J. Kleinberg, Signed Networks in Social Media. 
28th ACM Conference on Human Factors in Computing Systems (CHI), 1361-1370 (2010).
\bibitem{snap_url} https://snap.stanford.edu/data/
\bibitem{folena} G. Folena, G. Biroli, P. Charbonneau, Y. Hu, and F. Zamponi, 
	Equilibrium fluctuations in mean-field disordered models, arXiv:2202.07560 (2022).
\end{thebibliography}
\end{document}